\documentclass[11pt]{article}
\usepackage[margin=1in]{geometry}
\usepackage{graphics, array}
\usepackage{multirow}
\usepackage{natbib}
\usepackage{pslatex}
\usepackage{paralist}
\usepackage{amsmath, amsthm, amssymb} 
\begin{document}

\title{\sc A Top-down Model for Cash CLO}
\author{
Yadong Li and Ziyu Zheng
\thanks{Li: Barclays Capital, yadong.li@gmail.com. 
Zheng: Morgan Stanley, ziyu.zheng@ms.com.
This paper is based the authors' research between May and Sep 2008 while 
both were employed by Lehman Brothers. The views expressed in this paper 
are the authors' own and do not necessarily reflect those of Lehman Brothers or their 
current employers. The authors thank Greg Xue for Intex analytics
support, Lorraine Fan, Marco Naldi, Ariye Shater, Gaurav Tejwani 
for many helpful comments and discussions, and the Lehman CLO 
trading desk for many valuable inputs.
}
}

\maketitle

\begin{abstract}
We propose a top-down model for cash CLO. This model
can consistently price cash CLO tranches both within the same
deal and across different deals. Meaningful risk measures for cash 
CLO tranches can also be defined and computed. This method is
self-consistent, easy to implement and computationally efficient. 
It has the potential to bring much pricing transparency 
to the cash CLO markets; and it could also greatly improve the 
risk management of cash instruments.

\end{abstract}

\section{Introduction}

Throughout the credit crisis since 2007, cash structured finance 
instruments such as cash CDO, CLO, MBS and CMBS have suffered much larger 
losses and write-downs than the synthetic instruments such as synthetic
CDO/CLOs.  The cause of the disparity in losses is mainly due to the 
difference in their risk management practices. In the synthetic CDO/CLO 
market, the market participants are much more accustomed to hedging the
market risks via credit indices, index tranches and single name 
CDS/LCDS contracts. The base correlation model is the market standard 
model for the synthetic CDO/CLOs, which have played a key role in the 
risk management of synthetic CDO/CLOs. Having a market standard model 
like base correlation has nurtured and encouraged proper and sophisticated 
risk management practice across both investors and market makers in
the synthetic market. In comparison, the risk management in the cash 
CDO/CLO market is much less sophisticated, which is mainly due to the 
limited availability of hedging instruments as well as
the lack of a standard risk-neutral model. The model development 
for cash instruments has lagged far behind comparing to the modeling 
capabilities of the synthetic CDO/CLOs. The modeling
efforts of cash instruments have been mostly focused on the cashflow 
modeling, and there are few attempts to price cash CDO/CLOs 
under the proper risk-neutral framework. Risk management 
of cash instruments without a proper risk-neutral 
model is inevitably difficult.

In this paper, we present a simple and practical risk-neutral model 
for cash CLO (Collateralized loan obligation), which is one of the most 
important type of cash CDOs. The total outstanding 
notional amount of cash CLO assets has grown to over \$400B in 2007 right before 
the credit crisis. The same methodology can also be applied to other 
types of cash CDOs. The proposed methodology brings the modeling 
capability of cash instruments much closer to the modeling 
capability of the synthetic instruments. 

This paper is organized as follows, we first give a brief overview of 
the important differences between the cash and synthetic CLO markets 
in section \ref{intro}, then we review the existing pricing method for 
cash CLOs and its limitations in section \ref{existing}. In section 
\ref{topdown}, we introduce the top-down method for CLOs. In the
rest of this paper, the unqualified term ``CLO'' always refers to
the cash CLO, while the ``synthetic CLO'' is always fully qualified
to avoid confusion.

\section{Cash vs. Synthetic CLO\label{intro}}
A cash CLO and a synthetic CLO share very little in common besides their names. 
A cash CLO deal is usually managed by a manager who can actively 
trade and hedge the underlying loan portfolio, while a synthetic CLO
usually trades in single tranche format referencing a static portfolio. 
A cash CLO often has complex cashflow waterfall structure so that the cashflow 
of its tranches are highly path-dependent, while the cashflows 
of synthetic CLO tranche are very simple and not 
path-dependent. The cash CLO typically attracts institutional buy-and-hold 
 investors whose investment decisions are mainly based
on ratings and yields, while the synthetic CLO index (i.e.: the LCDX) 
and tranches are mostly traded by the correlation desks of hedge 
funds and banks who are required to mark their positions to the 
market. 

The market dynamics are totally different between the cash and 
synthetic CLO. In the synthetic CLO market, the participants 
can easily take long or short positions in the credit
indices (e.g. LCDX), index tranches and the underlying single name 
LCDS, therefore there are strong arbitrage relationships among them. 
It is easy to construct a profitable basis trade if any of the basis 
becomes very large. 
The basis between indices, tranches or single names in the 
synthetic market therefore tends to stay within a 
reasonable range. It is important for a synthetic CDO model 
to maintain the consistency between the tranches and its underlying 
portfolio because of this strong arbitrage relationship. The 
cashflows of synthetic CLO tranches are very simple and 
not path-dependent; therefore it is not only necessary but also
feasible to model synthetic CLOs as derivatives of the 
underlying LCDS contracts. This approach is commonly called ``bottom-up'' 
because the model drills down to the individual constituents 
in the collateral portfolio. Base correlation is the most common 
synthetic CDO/CLOs model in practice, readers are referred to 
\cite{bcexplained} for a description of the base correlation model. 
By modeling the synthetic CDO/CLO tranches as derivatives of the 
underlying CDS/LCDS, the total protection value of the tranches 
automatically adds up to the total protection value of the underlying 
CDS/LCDS.
 
In contrast, the cash CLOs behave very differently. 
A cash CLO is very similar to a regular company as it has a 
manager, some assets (the loan collateral) that are funded with 
various classes of debt (tranches) and equity (tranche).
Unlike the synthetic market, it is almost impossible to
construct a basis trade for cash CLOs even if there exists
large basis between the market value of CLO assets (underlying
collateral loans) and CLO liabilities (including tranches,
management fees and other expenses). There are several reasons 
for this: 1) cash CLO managers typically only report the 
underlying loan positions once a month, thus the exact loan 
positions are unknown between reports since the manager may
trade the underlying loans at any time 2) the individual
loans are often illiquid and rarely traded 3) taking short 
positions in underlying loans or cash CLO tranches is very 
difficult in the cash market. Therefore the basis trades for 
cash CLOs have rarely (if ever) been attempted in practice. 
As a result, there is no market force that brings the values 
of cash CLO assets and liabilities together quickly. The 
values of a cash CLO's assets and liabilities can diverge 
significantly for an extended time even though they eventually 
have to  converge\footnote{At the maturity of the CLO deal 
to the latest since all the loans will either mature or 
liquidate at maturity.}.  The cash CLO tranches and 
the underlying loans can behave like unrelated instruments in short 
term even though the cashflows of cash CLO tranches are derived from 
the cashflows of the underlying loans. It is not uncommon for the underlying 
loan prices and the cash CLO tranche prices to move in opposite 
direction during the same trading session. 

The cash CLO market dynamics therefore impose certain practical 
restrictions on the hedging strategies of cash CLO tranches:
\begin{enumerate}
\item
It is impractical and almost useless to hedge cash CLO tranches 
by trading individual loans. 
\item
It could be feasible and cost effective to macro hedge 
the overall market movements of the underlying loans using 
instruments such as TRS or LCDX index. The macro-hedges may 
not work well if the basis between CLO tranches and the 
underlying loan changes significantly, therefore one must 
have a view on the basis before putting on a macro hedge. 
\item
Even though the synthetic LCDX tranches were originally 
created to hedge the cash CLO tranches, the historical price 
movements of cash CLO tranches and synthetic LCDX tranches have 
shown very little correlation due to the fundamental differences 
in market dynamics. Therefore, the LCDX tranches have 
been very poor hedging instruments for the cash CLO tranches 
despite its original intention. 
\end{enumerate}
These practical hedging restrictions are important considerations
when building a cash CLO model.

\section{Review of Cash CLO Pricing Method\label{existing}}
Even though the cash CLO tranches trade in the 
upfront price format just like ordinary corporate bonds, a 
more structured quoting/pricing convention is required for cash CLO investors
to compare relative values among different cash CLO tranches. The current market 
standard for quoting and pricing cash CLO tranches is 
based on a single pricing scenario where the overall underlying 
loan collateral is assumed to have a constant annualized default
rate (CADR), constant annualized prepayment rate (CAPR), and a
constant recovery rate (CRR) over the whole life of the CLO.
The CADR of the single pricing scenario is usually very low,
for example 3\% CADR is often used. 
The CLO tranche cashflows from this single pricing scenario of
CADR, CAPR and CRR are then discounted using the riskfree rate 
plus a discount margin (DM) spread to produce the tranche PVs. 
Different DMs are required for different tranches to re-produce 
their market prices from the single pricing scenario. 

Intex\footnote{A product of Intex Solutions, http://www.intex.com} 
is a standard software package used by most market participants 
to compute the cashflows from CLO tranches. 
Intex has modeled the majority of outstanding cash CLO deals 
in the market and it can compute the cashflows of (almost) 
any CLO tranches under any CADR, CAPR and CRR scenarios. The wide adoption 
of Intex tool is important for the transparency of the cash CLO market 
because it provides a consensus on the CLO tranche 
cashflow among market participants. Different market participants
thus can reach the same conversion between the DM and 
CLO tranche prices as long as they all use Intex for the cashflow
calculation. The implied DM from the market tranche 
prices can then be used by the investors to compare the relative 
values of different cash CLO tranches. We can view the DM as a similar 
measure as the credit spreads for corporate bonds: the higher the DM, 
the more likely the cashflow from the single low-CADR pricing scenario 
would not be paid due to the increase in loan default rates. 

The cash CLO market makers often maintain a matrix of DMs based on recent 
market transactions of various ratings, vintages and deal types; and 
the DM matrix is used to price similar cash CLO tranches whose prices are not
observable in the market. Due to the lack of liquidity, the DM matrix 
is only updated infrequently, often once a month. This DM method is 
widely used because of its simplicity, but it is really a quoting 
convention rather than a pricing model. The following is a list of limitations
of the DM method: 
\begin{enumerate}
\item Some important cash CLO structural features, such
as IC/OC triggers, are not priced in since they are not relevant 
under the single low-CADR pricing scenario. However in reality, these 
features provide valuable protection to the senior and mezzanine tranches 
under the high default rate environment, therefore they should affect 
CLO tranche pricing.  
\item There is no pricing consistency across different tranches of the 
same cash CLO deal since their cashflows are discounted by different DMs. 
There is no concept of correlation of loan defaults and value shift between 
different parts of the CLO capital structure. 
\item The collateral loan prices do not enter the cash CLO tranche pricing
at all. Even though the values of cash CLO assets and liabilities do not 
necessarily move together as discussed in section \ref{intro}, the underlying 
loan prices is a key piece of information for comparing the relative values 
between different cash CLO deals since it reflects the expected 
loan default rates. 
\item There is no meaningful risk measures from the DM method
as the only risk factor of a cash CLO tranche is its DM by construction, 
which does not provide any useful guidance for risk management
and hedging. Simple questions like: ``how much a cash CLO tranche 
price would change if the underlying loan prices move by 1 point?''
cannot be answered.
\item The DM matrix is usually hand marked by the trading desk,
who can also adjust the DM of individual deals in the book. There is 
very little control in place on how a DM can be marked due to the 
illiquid nature of the cash CLO market. Often the tranches from certain 
CLO types and vintages do not transact in the market for weeks (or even
months). The DM marks and the resulting CLO tranche prices often
lack consistency and transparency; and they are easy targets 
for mis-marks and manipulations.
\end{enumerate} 

A new modeling paradigm is therefore urgently needed for
the cash CLO market. In this paper, we propose a top-down model
for CLO that addresses all the limitations above. 

\section{A Top-down Cash CLO Model\label{topdown}}
It is more difficult to construct proper models for cash CLOs
than for synthetic CLOs due to the opaqueness of the underlying 
collateral and the complex cashflow waterfall features.  
There have been prior attempts to price cash CLOs 
tranches by computing their cashflows from a Monte Carlo simulation
of the default times and recovery rates of underlying loans. 
The default time and recovery simulation could be driven
by a default time copula (e.g.: Gaussian Copula).
This bottom-up approach achieved little success because of the 
uncertainties in the underlying loan positions and prices, 
as well as the prohibitive computational cost to obtain the 
tranche cashflows from a large number of simulated scenarios. 
Custom implementations of the cash CLO's cashflow 
waterfalls are often required to achieve reasonable 
simulation speeds as the standard Intex tool may not be fast 
enough to support a large number of simulated scenarios.
In our view, it is not only a costly, but also an almost 
useless exercise to build a bottom-up cash CLO model in
practice. The main benefit of a bottom-up model is the ability
to produce risks to individual underlying loans, but it 
is not feasible to hedge cash CLO tranches by trading 
individual underlying loans anyway (as discussed in 
section \ref{intro}).

Recognizing the drawbacks of the current DM based method and 
the practical hedging restrictions in the cash CLO market, we 
hereby propose a simple but practical top-down model for cash 
CLO. Top-down models were originally developed for exotic 
synthetic instruments. In a top-down model, the collateral 
portfolio is modeled as a whole instead of drilling down to individual 
constituents. The benefit of a top-down approach 
is its simplicity as a result of not having to model the 
individual constituents of the underlying portfolio. The 
adoption of top-down models in synthetic CDO/CLOs has been 
limited so far because the prices of synthetic tranches do
move with its underlying CDS/LCDS due to the strong arbitrage 
relationship. Ignoring the single name risk is considered a 
drawback for synthetic instruments since it is critical to
hedge synthetic tranches by trading the underlying single 
name CDS/LCDS contracts. 

However, a top-down approach is ideal for cash CLOs 
since there is no strong arbitrage relationship between 
the CLO assets and liabilities in the cash market, and the cash 
CLO tranche prices do not necessarily move consistently with 
individual underlying loans. Therefore, 
ignoring the individual single name information, being a 
vice in the synthetic CLOs modeling, becomes a virtue 
in the cash CLO modeling because it is closer to the market
reality and it greatly simplifies the model setup. Note 
that a top-down model can produce risk measures to the 
overall average loan price in the cash CLO portfolio, 
thus it is possible to macro-hedge the overall loan price 
movements using a top-down model. It is just not possible 
to produce hedge ratios to the individual loans with a 
top-down approach, which is useless in practice anyway.

\subsection{Calibrate to ``Index'' CLO}
Fundamentally, the objective of a pricing model is to find the
prices of less liquid instruments from the prices of liquid 
instruments. For example, when pricing a bespoke synthetic CDO using the base 
correlation model, we first extracts the correlation information 
from the liquid index tranches by calibrating a base correlation 
surface, which is then mapped to the bespoke portfolios to produce the bespoke 
tranche prices. The pricing consistency among different bespoke 
tranches are maintained because all of them are priced from the 
same set of liquid index tranches. This procedure allows us to 
compute risk sensitivities of bespoke tranches to the liquid index 
tranches; and we can hedge the illiquid bespoke tranches 
by trading the liquid index tranches. 

Given the success of synthetic CDO/CLO models, it is a natural 
idea trying to apply the same pricing method to cash CLOs. 
However, a practical challenge is that there is no standard liquid 
index for cash CLOs. As discussed before, the LCDX tranches 
can't be used to price cash CLOs because of the fundamental differences 
between the cash and synthetic markets. To get around this, 
we have to assume that there is a representative cash CLO deal 
whose market price is somewhat transparent, which can be 
used as an ``index'' to price other cash CLO deals. In practice, 
cash CLO market participants can choose a representative 
CLO that are reasonably liquid as the CLO ``index''.
 Once we identified a cash CLO index and its tranche 
prices, we then can carry out the calibration and mapping 
procedure for cash CLOs in a similar manner as in the synthetic 
CLO models. Figure \ref{idx1} showed the deal information 
and tranche prices of an actual cash CLO deal, whose price marks 
are provided by the Lehman CLO trading desk as of Aug. 12, 2008. 
We will use this CLO deal as the ``index'' for the following 
discussion. This ``index'' deal is subsequently referred as
CLO-IDX.

\begin{figure}
\caption{CLO-IDX Deal Information\label{idx1}}
\center
\footnotesize

\begin{tabular}{c|crccc|r}
\hline
{\bf Class} & {\bf Coupon Rate} & {\bf Notional} & {\bf OC Trigger(\%)} & {\bf IC Trigger(\%)} & {\bf S\&P Rating} & {\bf Prices (\%)} \\
\hline
A     & Libor+69.5bp &   506,250,000  & \multirow{2}{*}{118.8} & \multirow{2}{*}{120.0} & AAA   & 92.97 \\
B     & Libor+110.0bp &     61,875,000  & &  & AA    & 82.16 \\
C     & Libor+200.0bp &     43,125,000  & 111.2 & 112.5 & A     & 78.83 \\
D     & Libor+3.25\% &     30,000,000  & 107.2 & 107.5 & BBB   & 72.13 \\
E     & Libor+5.00\% &     33,750,000  & 104.4 & 100.1 & BB    & 63.77 \\
SUBORD & -     &     75,000,000  & -     & -     & NA    & 44.89 \\
\hline
\end{tabular}

\end{figure}

The traditional DM method only uses a single pricing scenario 
of CADR, CAPR and CRR for all the tranches, which is an overly 
simplified assumption since the future default, prepay and
recovery rates are by no means deterministic. It is much more 
realistic to assume that there are a set of possible market 
scenarios of (CADR, CAPR and CRR), and each scenario has 
certain risk-neutral probability of realization. Figure 
\ref{cadrs} is a set of 
representative scenarios that are provided by the Lehman CLO 
research based on the market condition as of mid 2008.  
The CAPR and CRR are chosen to be decreasing with the CADR based on historical 
observations.  Even though the default, prepay and recovery 
rates can be time dependent, we kept them constant in this 
study for simplicity. It is sensible to add time varying 
default scenarios if default is likely to be front or back 
loaded. The distribution of these market 
scenarios can be calibrated to the market prices of the cash CLO 
tranches.
 
More formally, we use $S_i$ to represent the i-th (CADR, CAPR, 
CRR) scenario in Figure \ref{cadrs}, and we use $v_j(S_i)$ to 
represent the PV of the j-th CLO tranche under the 
scenario $S_i$, which can be computed by simply discounting 
the cashflow from Intex using the risk free rates {\bf without} 
any additional DM. We also use $V_j$ to represent
the market price of the j-th CLO tranche as shown in Figure 
\ref{idx1}, then the calibration reduces to a problem of
finding a discrete distribution of $\{p_i\}$ for the given 
set of scenarios so that for every CLO tranche $j$:
\begin{equation}
\label{calib}
\sum_i p_i v_j(S_i) = V_j
\end{equation}
We call the $\{p_i\}$ that solves \eqref{calib} the market implied scenario
distribution (MISD). This approach is similar in spirit to the \cite{brigogpl} 
for the synthetic CDOs, the main difference here is that there is no constraints 
from the underlying collateral loan prices, which is a conscious choice 
because there is no strong arbitrage relationship between the cash 
CLO assets and liabilities. The tranche prices across the full capital 
structures have to be used in the MISD calibration otherwise the overall 
risk of the underlying portfolio cannot be determined. In this approach, 
the CLO tranche cashflows are computed by Intex using only the aggregated 
CADR, CAPR and CRR 
of the whole loan collateral portfolio, this is effectively a top-down 
approach since it does not drill down to the individual collateral
loans.

\begin{figure}
\caption{Loan Market Scenarios\label{cadrs}}
\center
\footnotesize

\begin{tabular}{rrr|rrr}
\hline
{\bf CADR (\%)} & {\bf CAPR (\%)} & {\bf CRR (\%)} & {\bf CADR (\%)} & {\bf CAPR (\%)} & {\bf CRR (\%)} \\
\hline
0     & 15    & 84    & 16    & 0     & 36 \\
1     & 14    & 81    & 17    & 0     & 33 \\
2     & 13    & 78    & 18    & 0     & 30 \\
3     & 12    & 75    & 19    & 0     & 27 \\
4     & 11    & 72    & 20    & 0     & 24 \\
5     & 10    & 69    & 22    & 0     & 18 \\
6     & 9     & 66    & 24    & 0     & 12 \\
7     & 8     & 63    & 26    & 0     & 6 \\
8     & 7     & 60    & 28    & 0     & 0 \\
9     & 6     & 57    & 30    & 0     & 0 \\
10    & 5     & 54    & 35    & 0     & 0 \\
11    & 4     & 51    & 40    & 0     & 0 \\
12    & 3     & 48    & 45    & 0     & 0 \\
13    & 2     & 45    & 50    & 0     & 0 \\
14    & 1     & 42    & 60    & 0     & 0 \\
15    & 0     & 39    & 90    & 0     & 0 \\
\hline
\end{tabular}

\end{figure}

Since the number of scenarios in Figure \ref{cadrs} is much
greater than the number of tranches in Figure \ref{idx1}, there
are infinitely many distributions that can reprice all
the index CLO tranches. Therefore, certain objective function 
has to be exogenously chosen so that we can find a unique distribution 
using an optimization method. The maximum entropy method 
is well suited for such under-determined optimization 
problems in derivative pricing as it finds a distribution with 
the most uncertainty and the least bias. Readers are referred 
to \cite{wmc} for an introduction to the maximum entropy
optimization method.

\begin{figure}
\caption{CLO-IDX Tranche Prices under Market Scenario \label{intex}}
\center
\footnotesize

\begin{tabular}{r|rrrrrr|r}
\hline
{\bf CADR} & {\bf  A} & {\bf B} & {\bf C} & {\bf D} & {\bf E} & {\bf SUB} & {\bf COL} \\
\hline
0     & 103.92 & 107.69 & 114.67 & 124.66 & 139.01 & 174.62 & 109.73 \\
1     & 103.93 & 107.74 & 114.81 & 124.83 & 139.49 & 162.27 & 109.07 \\
2     & 103.95 & 107.83 & 114.95 & 125.15 & 139.81 & 147.21 & 108.22 \\
3     & 103.97 & 107.91 & 115.16 & 125.53 & 140.37 & 127.51 & 107.18 \\
4     & 103.99 & 108.03 & 115.38 & 125.89 & 141.12 & 101.18 & 105.93 \\
5     & 104.01 & 108.09 & 115.52 & 126.12 & 141.33 & 73.03 & 104.49 \\
6     & 104.03 & 108.23 & 115.80 & 126.64 & 136.62 & 43.27 & 102.86 \\
7     & 103.96 & 108.21 & 115.88 & 126.86 & 129.57 & 17.97 & 100.98 \\
8     & 103.46 & 107.61 & 115.77 & 127.04 & 112.86 & 13.46 & 98.94 \\
9     & 102.86 & 106.23 & 112.14 & 125.50 & 80.36 & 10.26 & 96.66 \\
10    & 102.68 & 105.95 & 111.33 & 120.69 & 35.28 & 7.12  & 93.99 \\
11    & 102.62 & 105.98 & 111.63 & 61.10 & 29.66 & 6.22  & 91.33 \\
12    & 102.57 & 105.99 & 105.77 & 8.91  & 25.43 & 4.30  & 88.48 \\
13    & 102.54 & 106.15 & 62.05 & 7.26  & 18.14 & 3.16  & 85.44 \\
14    & 102.61 & 105.57 & 5.81  & 5.67  & 18.55 & 3.09  & 82.21 \\
15    & 102.61 & 70.31 & 5.80  & 5.67  & 13.53 & 3.01  & 78.78 \\
16    & 102.07 & 36.85 & 4.53  & 4.11  & 13.29 & 2.94  & 75.67 \\
17    & 98.08 & 33.64 & 3.28  & 4.11  & 8.68  & 2.26  & 72.50 \\
18    & 93.76 & 32.90 & 3.28  & 4.11  & 12.42 & 2.10  & 69.28 \\
19    & 89.02 & 32.90 & 3.28  & 2.58  & 12.31 & 0.31  & 66.02 \\
20    & 84.03 & 32.90 & 2.40  & 2.58  & 12.14 & 0.20  & 62.72 \\
22    & 73.89 & 31.80 & 2.05  & 2.58  & 11.80 & 0.16  & 56.02 \\
24    & 63.67 & 31.74 & 2.05  & 2.58  & 7.60  & 0.13  & 49.22 \\
26    & 54.89 & 29.79 & 2.05  & 2.58  & 1.31  & 0.09  & 42.35 \\
28    & 44.48 & 29.38 & 2.05  & 1.02  & 2.39  & 0.00  & 35.43 \\
30    & 40.43 & 29.39 & 0.80  & 1.02  & 9.22  & 0.00  & 32.76 \\
35    & 33.23 & 17.47 & 0.80  & 1.02  & 8.78  & 0.00  & 27.08 \\
40    & 27.53 & 10.79 & 0.80  & 1.02  & 3.70  & 0.00  & 22.60 \\
45    & 36.50 & 7.98  & 0.80  & 1.02  & 0.00  & 0.00  & 19.09 \\
50    & 33.40 & 4.77  & 0.00  & 0.00  & 0.00  & 0.00  & 16.32 \\
60    & 28.28 & 3.71  & 0.00  & 0.00  & 0.00  & 0.00  & 12.39 \\
90    & 20.74 & 1.68  & 0.00  & 0.00  & 0.00  & 0.00  & 6.55 \\
\hline
\end{tabular}

Market scenarios are only indexed by its CADR, the full scenario
definition is listed in Figure \ref{cadrs}.
\end{figure}

\begin{figure}
\caption{Calibrated MISD\label{dist0}}
\vspace{.25cm}

\center
\begin{minipage}{3in}
\center
\underline{CLO-IDX}
\scalebox{.55}{\includegraphics{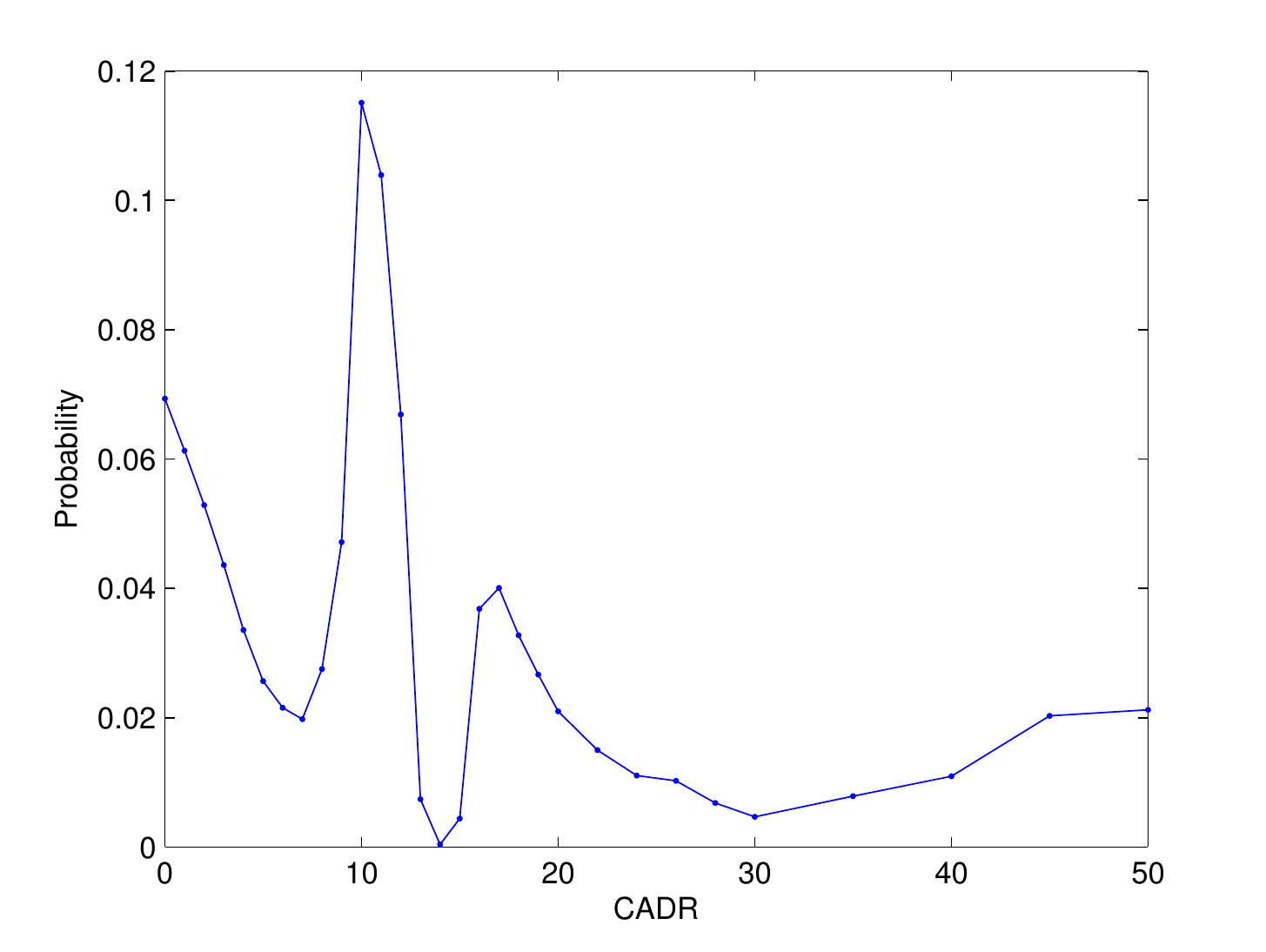}}
\end{minipage}
\begin{minipage}{3in}
\center
\underline{LCDX10}
\scalebox{.55}{\includegraphics{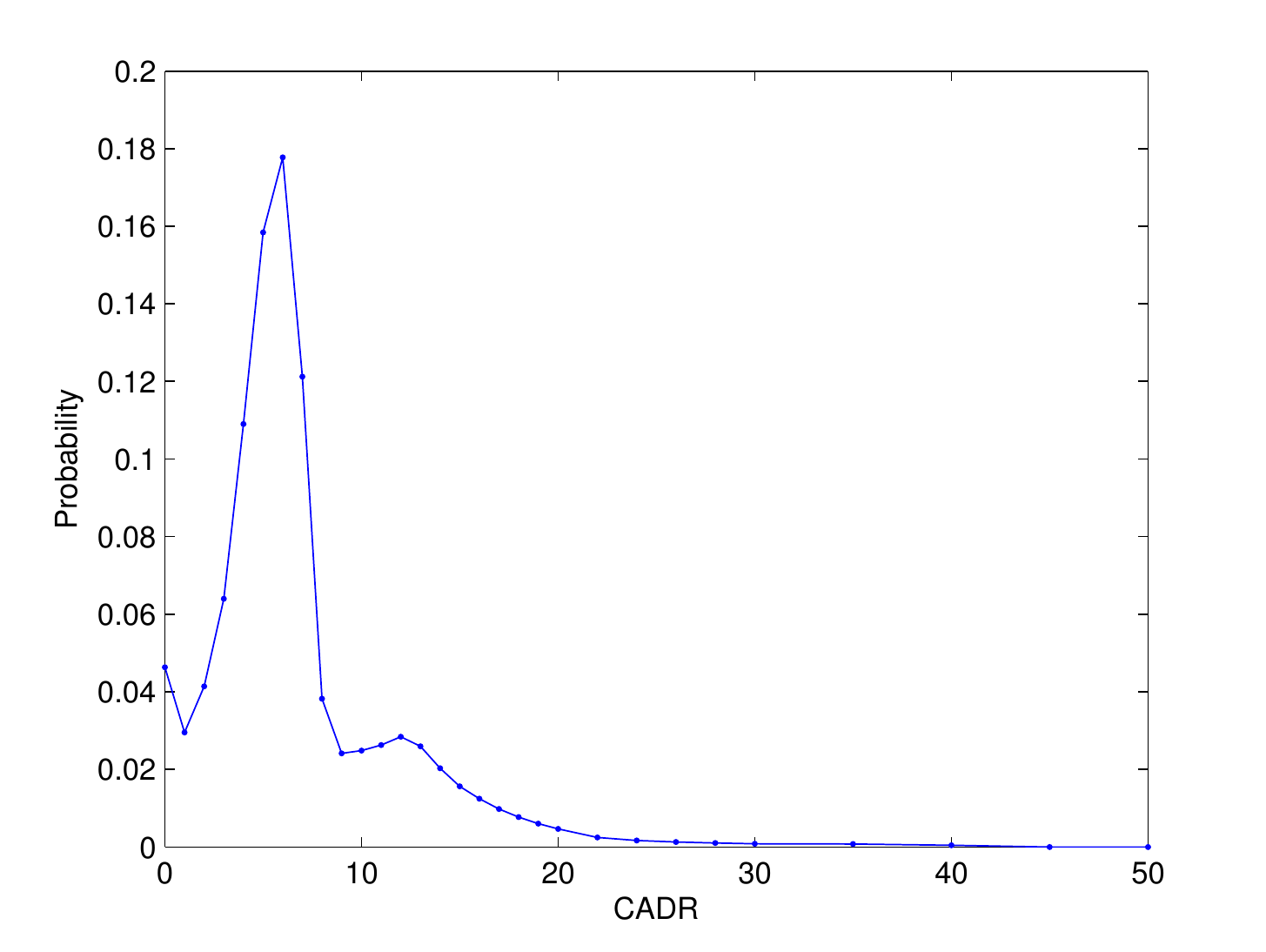}}
\end{minipage}

\end{figure}

Figure \ref{intex} showed the CLO-IDX tranche prices computed by
Intex for every scenario in Figure \ref{cadrs}. Note that these 
tranche prices are computed {\bf without} any additional DM
above the risk free rate.  Given 
the data in Figure \ref{intex}, we can easily find the MISD that 
reproduces the market CLO tranche prices in Figure \ref{idx1} 
via the maximum entropy method. The calibrated 
distribution is shown in the left side of Figure \ref{dist0}. 
The fitting quality of the MISD is excellent, the market tranche 
prices are matched almost exactly, which is not surprising 
because the number of tranches is much less than the number 
of market scenarios. 

By finding the MISD, we have moved from the traditional pricing 
method of a single pricing scenario with different DMs for different 
tranches, to a more consistent pricing method of a single MISD and 
a single set of risk-free discount factors for all tranches. It seems to be a
small step to replace multiple DMs with multiple market scenarios
in the MISD, however this is a significant step forward since 
it is not only more realistic, but also addresses the first 
two limitations of the traditional DM method listed in section 
\ref{existing}. Since the market scenarios in Figure \ref{cadrs} 
cover a wide range of CADR from 0\% to 90\%, every structural feature 
in a cash CLO deal is expected to be triggered under some of
the scenarios; thus they are fully priced in by the MISD 
method. Also, it is obvious that all the tranches from 
the same CLO deal are priced consistently to each other because 
the same MISD and risk-free discount factors are used. 

\begin{figure}
\caption{CLO-IDX Calibration with Bumped AA Tranche Prices\label{aa}}
\center
\scalebox{.55}{\includegraphics{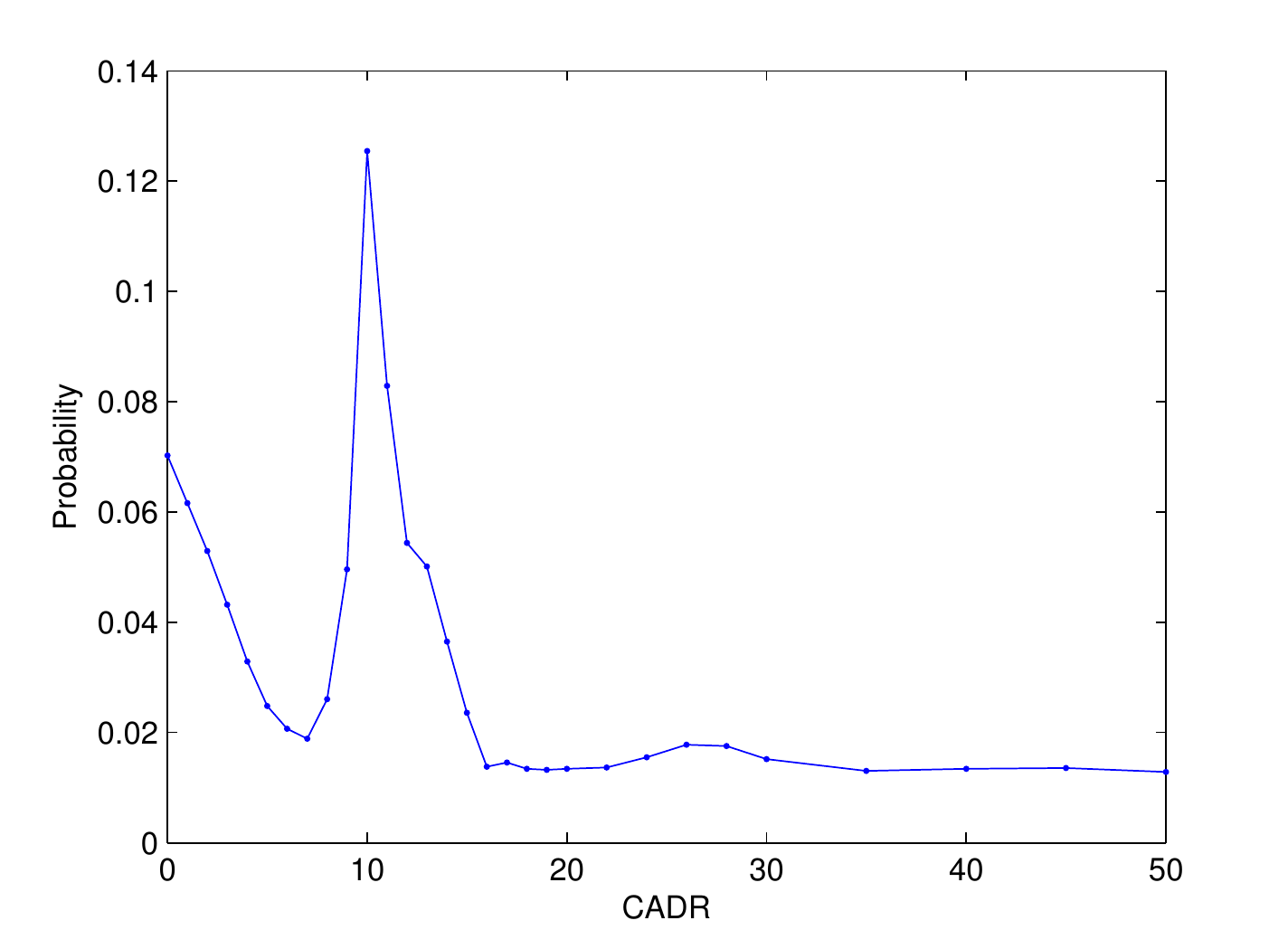}}
\end{figure}

For comparison purposes, we also calibrated the MISD to the
synthetic LCDX10 tranches, the results are shown on the right side of 
Figure \ref{dist0}. The tranche PVs of LCDX10 under each market 
scenario can be directly computed without using Intex 
since the LCDX tranche cashflow is a simple function of the 
aggregated portfolio loss. At first glance, the MISD from
the CLO-IDX and the LCDX10 are quite different: the MISD 
from LCDX10 is roughly uni-modal while the MISD from CLO-IDX is
obviously multi-modal. We have calibrated the MISD to many cash CLO
deals and found that the multi-modality of MISD is a common 
feature among almost all cash CLOs in mid 2008, whereas it is 
not present in the MISD from synthetic CLOs such as LCDX10. 
The multi-modality of CLOs is caused by the strong market 
demand for the very safe 
AAA rated assets during the severe market stress of mid 2008. 
As shown in Figure \ref{idx1}, the AA tranche is priced more 
than 10 points cheaper than the AAA tranche, such a steep price 
drop from the AAA tranche to AA tranche is mostly caused
by market technicals instead of fundamentals. If we bump
the AA tranche price in Figure \ref{idx1} up by 5 points 
and re-calibrate the MISD to the bumped CLO prices,
the resulting MISD (Figure \ref{aa}) becomes closer to the 
uni-modal MISD from the LCDX10. This exercise showed that the 
MISD method allows us to meaningfully compare and identify 
discrepancies between cash and synthetic CLO markets, it also 
demonstrated the huge differences in dynamics and technicalities 
between cash and synthetic CLO markets. 

\begin{figure}
\caption{Tranche Implied Expectations for CLO-IDX\label{implied}}
\center
\footnotesize

\begin{tabular}{c|r}
\hline
{\bf CADR}  & 13.94\% \\
{\bf CAPR} & 5.77\% \\
{\bf CRR} & 49.94\% \\
{\bf Average Collateral Loan Price} & 84.42 \\
\hline
\end{tabular}
\end{figure}

From the calibrated MISD, we can easily compute expectations of 
various quantities, such as expected CADR, CAPR and CRR
rates, as shown in Figure \ref{implied}. Since the MISD are 
calibrated to cash CLO tranche prices, we call them 
the tranche implied CADR, CAPR and CRR. The tranche
implied CRR is much lower than the historical loan recovery 
rates (usually above 70\%), which is a sign of stress 
in the cash CLO tranche market. 

Using the average underlying collateral loan price (last
column of Figure \ref{intex}) computed by Intex, we can also obtain 
the tranche implied average collateral loan price, which is the average
collateral loan price that makes the total asset value equals the 
total liability 
value for the given cash CLO deal. Comparing the tranche implied average loan 
price against the average of actual market loan prices\footnote{
Since some of the underlying loans are illiquid, the average loan prices 
are actually computed only from loans with observable market
prices} gives us the basis between the cash CLO tranche market 
and the underlying loan market. This basis cannot be obtained by 
simply comparing the notional weighted average of the 
CLO tranche prices against the average price of the underlying
loans since 1) this does not take into account the cash CLO manager's 
fee and other expenses that are taken from the collateral loan 
cashflows 2) the total notional amount of the cash CLO tranches is usually 
not the same as the total notional amount of the underlying loans. 
The COL column in Figure \ref{intex}
reported by Intex, on the other hand, is the PV of the total 
underlying collateral loan (inclusive of any fees and expenses) 
normalized by the outstanding notional of the loan collateral,
therefore it is the right quantity for computing the tranche
implied average loan prices. 

In the case of CLO-IDX, the average market price of its underlying 
loans is around 89.51 and the tranche implied average collateral loan 
price is 84.42 as shown in Figure \ref{implied}, i.e.,
there is a very large negative basis of -5 points, which implies that
a investor would lose 5 points instantly if he creates a cash CLO out 
of a pool of loans. At the mid of 2008, almost all cash CLO 
deals showed large negative basis between the tranche 
implied loan price and the average market loan price. In a normal market 
environment, this basis should be positive otherwise there is no 
economic incentive to package individual loans to CLOs in the 
first place\footnote{Release of capital is another reason to
create cash CLO from loans.}. However, during market stress of mid 
2008, the cash CLO tranches are severely depressed amid the 
wide-spread fear of complex structured finance products, therefore 
the cash CLO tranches traded at deep discounts comparing to the 
underlying loans. 

It is another important advantage for the top-down MISD method 
to be able to compute the tranche implied quantities and compare 
them against those from the underlying loan market. This offers a 
meaningful relative value comparison between the cash CLO market 
and the underlying loan market. These tranche 
implied quantities cannot be computed from the traditional DM 
method, nor can they be obtained from any bottom-up models 
because bottom-up models enforces the value equality between
assets and liabilities. Therefore, the top-down MISD approach 
is actually more useful and closer to market reality than the 
bottom-up approach for cash CLOs.

\begin{figure}
\caption{CLO-BSPK Deal Information \label{bspkinfo}}
\center
\footnotesize

\begin{tabular}{c|crccc}
\hline
{\bf Class} & {\bf Coupon Rate} & {\bf Notional} & {\bf OC Trigger(\%)} & {\bf IC Trigger(\%)} & {\bf S\&P Rating} \\
\hline
A     & Libor+25.0bp & 375,000,000  & \multirow{2}{*}{111.7} & \multirow{2}{*}{111.7} & AAA \\
B     & Libor+37.0bp & 22,500,000  &  &  & AA \\
C     & Libor+65.0bp & 17,500,000  & 108.9 & 108.9 & A \\
D     & Libor+140.0bp & 30,000,000  & 103.5 & 103.5 & BBB \\
E     & Libor+3.650\% & 15,000,000  & 102.1 &    -   & BB \\
SUBORD & -     & 40,000,000  &    -  &   -   & NA \\
\hline
\end{tabular}
\end{figure}

\subsection{Map to ``Bespoke'' CLO}
After calibrating the MISD to an ``index'' CLO, we now investigate how to 
price tranches of other CLO deals. Borrowing the terminology from synthetic CLO, 
we refer the cash CLO deal we want to price as the ``bespoke'' CLO. In this section,
we use another actual CLO as a sample bespoke CLO, which we subsequently refer to as 
CLO-BSPK. The CLO-BSPK is of the same vintage and has similar structural features 
as the CLO-IDX, therefore it is quite sensible to price CLO-BSPK from CLO-IDX. 
Figure \ref{bspkinfo} shows some basic information on the CLO-BSPK deal, and 
Figure \ref{bspk12cf} shows the tranche prices of CLO-BSPK calculated by Intex 
under the market scenarios of Figure \ref{cadrs}.

\begin{figure}
\caption{CLO-BSPK Tranche Prices under Market Scenarios\label{bspk12cf}}
\center
\footnotesize

\begin{tabular}{r|rrrrrr|r}
\hline
{\bf CADR} & {\bf A} & {\bf B} & {\bf C} & {\bf D} & {\bf E} & {\bf SUB} & {\bf COL} \\
\hline
0     & 101.30 & 102.48 & 104.50 & 110.08 & 127.30 & 185.91 & 106.74 \\
1     & 101.31 & 102.50 & 104.54 & 110.20 & 127.55 & 167.50 & 105.97 \\
2     & 101.31 & 102.53 & 104.60 & 110.32 & 127.87 & 145.65 & 105.03 \\
3     & 101.32 & 102.56 & 104.67 & 110.48 & 128.21 & 120.74 & 103.92 \\
4     & 101.32 & 102.58 & 104.70 & 110.53 & 128.32 & 100.83 & 102.64 \\
5     & 101.33 & 102.62 & 104.79 & 110.69 & 122.26 & 71.86 & 101.19 \\
6     & 101.08 & 102.12 & 103.90 & 108.98 & 112.89 & 44.05 & 99.57 \\
7     & 100.98 & 101.99 & 103.60 & 108.20 & 111.50 & 17.45 & 97.78 \\
8     & 100.92 & 101.95 & 103.50 & 107.90 & 86.04 & 11.69 & 95.80 \\
9     & 100.90 & 101.94 & 103.49 & 96.37 & 39.64 & 9.07  & 93.66 \\
10    & 100.88 & 101.96 & 103.54 & 61.12 & 38.19 & 6.03  & 91.34 \\
11    & 100.86 & 101.98 & 102.66 & 23.67 & 37.34 & 3.57  & 88.85 \\
12    & 100.88 & 102.02 & 63.94 & 5.27  & 33.25 & 2.31  & 86.18 \\
13    & 100.86 & 92.13 & 6.31  & 4.39  & 32.66 & 2.29  & 83.34 \\
14    & 100.88 & 36.86 & 5.27  & 3.98  & 16.62 & 2.26  & 80.33 \\
15    & 97.40 & 27.70 & 4.25  & 2.88  & 16.37 & 2.24  & 77.14 \\
16    & 93.60 & 27.42 & 3.30  & 2.88  & 16.15 & 2.22  & 74.20 \\
17    & 89.72 & 27.42 & 3.30  & 2.88  & 15.84 & 2.20  & 71.21 \\
18    & 86.13 & 27.42 & 2.38  & 2.88  & 9.10  & 1.46  & 68.17 \\
19    & 82.26 & 27.42 & 2.38  & 1.80  & 9.33  & 0.62  & 65.08 \\
20    & 78.11 & 27.28 & 2.38  & 1.80  & 13.74 & 0.00  & 61.94 \\
22    & 69.84 & 26.39 & 1.49  & 1.80  & 15.52 & 0.00  & 55.56 \\
24    & 61.31 & 26.39 & 1.49  & 1.80  & 15.30 & 0.00  & 49.06 \\
26    & 52.80 & 25.34 & 1.49  & 0.70  & 15.07 & 0.00  & 42.46 \\
28    & 44.06 & 25.34 & 1.49  & 0.70  & 14.84 & 0.00  & 35.79 \\
30    & 40.70 & 24.28 & 1.49  & 0.70  & 14.60 & 0.00  & 33.03 \\
35    & 35.03 & 22.17 & 0.58  & 0.70  & 0.00  & 0.00  & 27.09 \\
40    & 29.53 & 10.57 & 0.58  & 0.70  & 0.00  & 0.00  & 22.28 \\
45    & 24.79 & 9.08  & 0.58  & 0.00  & 0.00  & 0.00  & 18.46 \\
50    & 21.15 & 4.87  & 0.58  & 0.00  & 0.00  & 0.00  & 15.44 \\
60    & 15.71 & 3.93  & 0.00  & 0.00  & 0.00  & 0.00  & 11.15 \\
90    & 8.48  & 2.19  & 0.00  & 0.00  & 0.00  & 0.00  & 5.38 \\
\hline
\end{tabular}

Market scenarios are only indexed by its CADR, the full scenario
definition is listed in Figure \ref{cadrs}.
\end{figure}

\begin{figure}
\center
\caption{Mapping from Index to Bespoke\label{bspk}}

\scalebox{.55}{\includegraphics{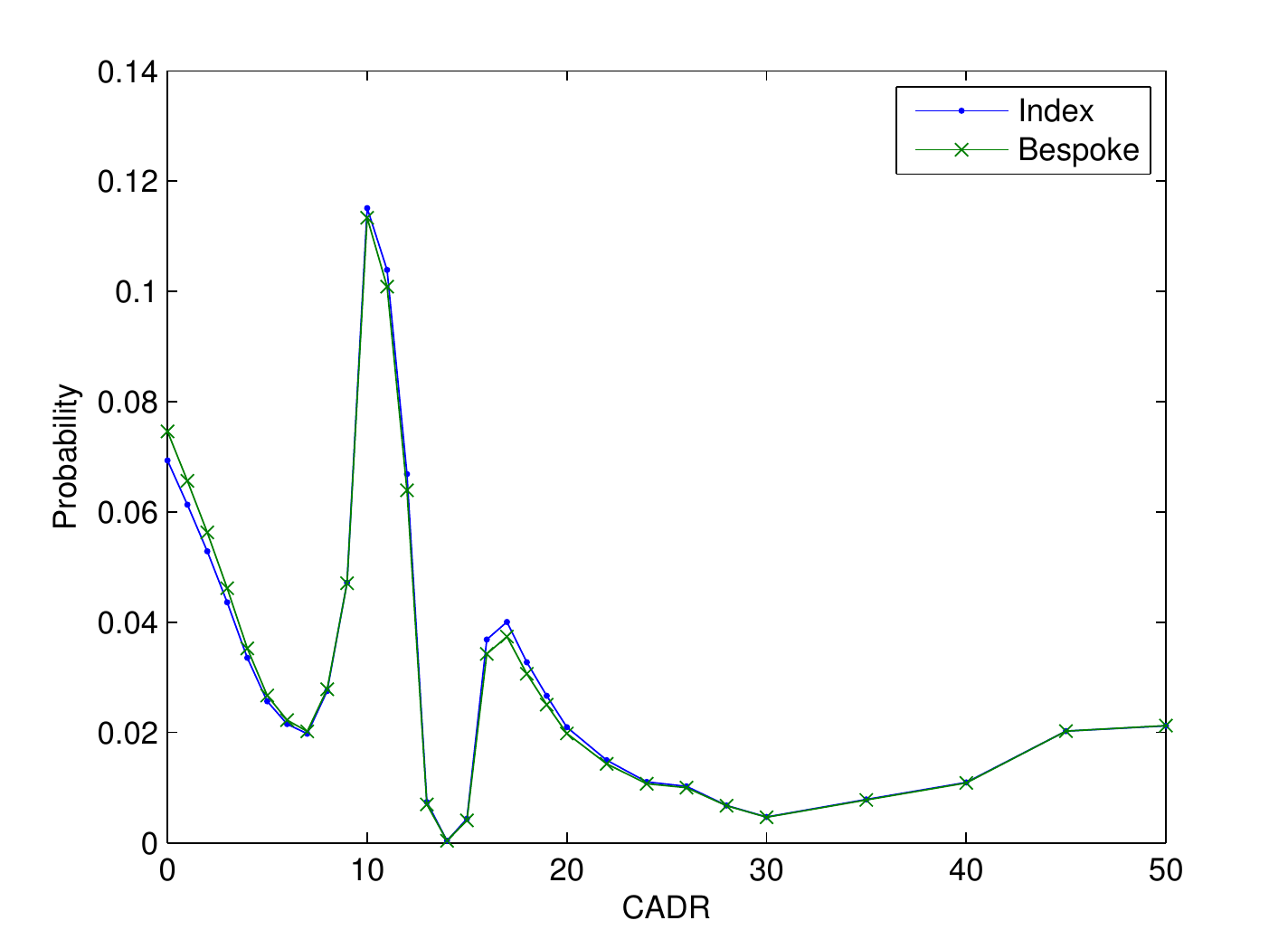}}
\end{figure}

The key to price a ``bespoke'' CLO is to perturb the index MISD to  
incorporate the bespoke specific information. Using terminology
from synthetic CLO, we need to find a mapping methodology 
between the index MISD and the bespoke MISD. The cross entropy 
method is an ideal method for such mapping operation between two
distributions since it gives a distribution that is closest to 
the prior distribution (i.e., the calibrated index MISD) while
satisfying additional linear constraints that accounts for 
important bespoke specific features. Readers are referred to 
\cite{wmc} for an introduction to the cross entropy (also
known as Kullback-Leibler relative entropy) method. Among all 
the features of the ``bespoke'' CLO, two of them
are the most important: 
\begin{enumerate}
\item The average price of the underlying loan collateral: This
is important because it adjusts for the loan quality difference 
between the index and bespoke CLO. It also allows us to 
compute tranche sensitivities to the average loan prices, which
can be used for macro-hedging.
\item The AAA-rated tranche: Since the AAA tranches are the 
most liquid, and all the AAA CLO tranches are priced very 
similarly to each other with minor adjustments for coupons and 
the underlying loan quality. The market participants can accurately
determine the bespoke AAA tranche prices from similar AAA 
transactions on the market.
\end{enumerate}
Both of these features can be easily added as linear constraints
in the cross entropy optimization. To incorporate the underlying
loan price, we need an assumption on the size of the basis
between the tranche implied and market loan prices for the 
bespoke CLO. Here we assume that the basis is constant and we 
use $K^I$ to denote the basis between the tranche implied loan 
price and average market loan price for the index CLO and 
$M^B$ to denote the average market price of the underlying loans 
in the bespoke portfolio, then the linear constraint for the 
market loan price of the bespoke CLO becomes:
\begin{equation}
\label{basis}
\sum_i q_i C(S_i) = K^I + M^B
\end{equation}
where $q_i$ is the MISD of the bespoke cash CLO we want to price; 
$C_j(S_i)$ is the average collateral loan prices from Intex as shown in the
last column of Figure \ref{bspk12cf}, which already accounts for
the average underlying loan features such as coupon rate, payment schedule 
and amortization etc.  Similarly, the constraints for the 
AAA tranche can be expressed as:
\begin{equation}
\label{aaa}
\sum_i q_i v^B_\textup{AAA}(S_i) = V^B_\textup{AAA}
\end{equation} 
where $v^B_\textup{AAA}$ is the tranche PVs for the AAA tranche in 
Figure \ref{bspk12cf}, and the $V^B_\textup{AAA}$ is the expected
bespoke AAA tranche price.

Other adjustments can also be included in the cross entropy 
mapping method. For example, CLO deals managed by a reputable 
manager often command a sizable premium comparing to those
managed by a mediocre manager. This management quality factor
can also be included as an adjustment to \eqref{basis}. In this 
example, we assume there is no difference in management quality 
between the CLO-BSPK and CLO-IDX.

\begin{figure}
\caption{Bespoke Cash CLO Pricing\label{bspkprice}}

\center
\footnotesize

\begin{tabular}{c|crr}
\hline
{\bf Class} & {\bf S\&P Rating} & {\bf Model Price} & {\bf Model Delta} \\
\hline
CLO-BSPK A & AAA   & 89.35 & 0.30 \\
CLO-BSPK B & AA    & 78.24 & 1.52 \\
CLO-BSPK C & A     & 70.78 & 2.38 \\
CLO-BSPK D & BBB   & 55.90 & 3.55 \\
CLO-BSPK E & BB    & 60.95 & 3.77 \\
CLO-BSPK SUBORD & -  & 47.63 & 5.21 \\
\hline
\end{tabular}

\end{figure}

With these two linear constraints in \eqref{basis} and \eqref{aaa}, 
it is easy to find the MISD for 
the bespoke cash CLO via the cross entropy method. Figure 
\ref{bspk} showed both the MISD of the index cash CLO and the 
mapped MISD of the bespoke CLO. The bespoke cash CLO tranche prices 
are easy to compute from the 
mapped MISD and the PV scenarios in Figure \ref{bspk12cf}. Figure 
\ref{bspkprice} showed the resulting bespoke cash CLO tranche prices. 

It is interesting to note that the E tranche is priced higher
than the D tranche for CLO-BSPK as shown in Figure \ref{bspkprice}. 
A careful examination of the tranche PVs in Figure \ref{bspk12cf} 
reveals the reason being that E tranche worths much more than the 
D tranche under most high CADR scenarios due to a structural 
features called CERT trigger in the CLO-BSPK deal, which diverts
the cashflow to the E tranche instead of the more senior C and D
tranches under certain high default rate scenarios. As shown in 
the cashflow table in Figure \ref{intex}, a similar but less 
prominent CERT trigger also exists in the CLO-IDX deal. The
purpose of the CERT trigger was to boost the rating of the E
tranche. If we use the traditional DM method to price the 
CLO-BSPK, the E tranche would certainly be priced less 
than the D tranche since the DM method only uses the single 
pricing scenario of 3\% CADR, under which the D and E tranches 
of CLO-BSPK behave very similarly to the D and E tranches of 
CLO-IDX since the CERT trigger is only active under high 
default rate scenarios. This example showed that the top-down 
MISD method automatically prices in the structural differences
between the index and bespoke cash CLO tranches, thus being able
to identify potential mis-prices in the traditional DM method. 

The proposed calibration and mapping procedure is a one-period
model without any term structure. Unlike synthetic CDO/CLOs which can 
trade at multiple maturities, each cash CLO only have a single 
pre-determined maturity, therefore a one-period model 
is adequate if the index and bespoke CLOs are from similar
vintage and have similar reinvestment period. 

\subsection{Risk Measures}
Under the traditional DM method, it is very difficult to quantify the
risk of a cash CLO book since there are no meaningful risk measures. The most
common view of the risk is the aggregated tranche notional for each 
rating bucket, which is a very crude estimate of the overall risk as 
there is no indication of the relative riskiness between different 
rating buckets. No concrete hedging strategy can be devised from the
aggregated cash CLO notional amounts by the rating bucket.

With the cross-entropy mapping method between the MISD of the index 
and bespoke CLO, we can easily define and compute a set of consistent
risk measures for the CLO tranches. For example, it is easy to 
compute the cash CLO tranche sensitivities to the underlying loan price 
movements via a simple bump-remap-reprice procedure. As shown in 
Figure \ref{bspkprice}, the tranche deltas computed this way are quite 
reasonable as the deltas are positive and decreasing with the tranche 
seniority. The aggregated tranche deltas can be used to predict the 
P\&L change of the whole book for a given movement of the average 
underlying loan price. This is a very precise risk measure which can be
used to macro-hedge the cash CLO book. 

\begin{figure}
\caption{CLO-BSPK Tranche01 Risk\label{tranche01}}

\center
\footnotesize

\begin{tabular}{c|rrrrrr}
\hline
{\bf Bespoke Tranche} & {\bf AAA} & {\bf AA} & {\bf A} & {\bf BBB} & {\bf BB} & {\bf NA} \\
\hline
CLO-BSPK A & {\bf 0.89} & {\bf 0.24} & -0.20 & 0.01  & -0.01 & 0.00 \\
CLO-BSPK B & 0.11  & {\bf 0.43} & {\bf 0.31} & 0.00  & -0.01 & 0.00 \\
CLO-BSPK C & 0.25  & -0.49 & {\bf 1.07} & {\bf 0.27} & -0.08 & 0.01 \\
CLO-BSPK D & -0.02 & -0.01 & -0.02 & {\bf 0.50} & {\bf 0.54} & -0.06 \\
CLO-BSPK E & 0.04  & 0.12  & 0.12  & -0.08 & {\bf 0.72} & 0.15 \\
CLO-BSPK SUBORD & 0.03  & -0.05 & 0.14  & 0.04  & -0.04 & {\bf 1.08} \\
\hline
\end{tabular}

\end{figure}

Similarly, we can define and compute the sensitivities to the 
index cash CLO tranches 
via the bump-remap-reprice procedure, as shown in Figure \ref{tranche01}. 
This sensitivity is commonly called tranche01 in the synthetic CLO 
terminology. The tranche01 risk of CLO-BSPK showed that the CLO-BSPK
tranches are the most sensitive to the index tranches of the 
same rating with some spillover to the next junior tranche. The tranche01 
risk is a measure of correlation risk, which is very useful in practice 
because we can break down the  risk of a cash CLO trading 
book into corresponding tranches of the index CLO, thus allowing us 
to understand and manage the risk exposure to different parts of 
the capital structure. The tranche01 risk is a much better measure
than the aggregated tranche notional by rating bucket. For example,
Figure \ref{tranche01} showed that the BBB-rated D tranche of the CLO-BSPK
behaves like a 50-50 mix of the D(rated BBB) and E(rated BB) 
tranches of the CLO-IDX; which is mainly due to the fact that the 
D tranche of CLO-BSPK has 3.5\% less subordination than the D tranche 
of CLO-IDX even though they are both BBB rated. These structural
differences between cash CLO deals are automatically captured by the
tranche01 risk from the top-down MISD method, thus providing a
much more coherent view of the correlation risk of a cash CLO book.

Other risk measures such as interest rate risk and theta risk can
be similarly defined and computed from the top-down method.

\section{Conclusion}
The proposed top-down method is ideal for cash CLOs 
as it produces consistent prices and risks across cash CLO 
deals while being very simple, intuitive and computationally 
efficient. Drilling down to the individual collateral loans provides 
very little practical benefits because of the lack of liquidity 
in individual loans and the lack of strong arbitrage relationship 
between cash CLO assets and liabilities.

In practice, this top-down model allows a cash CLO trading desk to 
only mark the prices of a few representative cash CLO deals as the 
indices for different vintages and deal types, then the rest 
of the cash CLO tranches in the book can be automatically priced 
via the cross entropy mapping method. This allows the cash CLO 
tranches to be priced consistently using the same calibrate-and-mapping 
procedure as in synthetic CDO/CLOs, making it much more difficult to 
manipulate the price marks and book P\&L. 

All the structural features of a cash CLO are automatically taken into 
consideration by the proposed top-down method, which is a big improvement 
over the traditional DM based method. Although the consistency between the 
value of cash CLO assets and liability is not enforced during the 
calibration because of the lack of strong arbitrage relationship, 
the average loan price is a valuable piece of market information 
and it is used by the cross entropy mapping to adjust for the underlying 
loan quality difference between cash CLO deals. Therefore this 
top-down MISD method can be a very effective method in finding 
relative value trading opportunities between cash CLO tranches, 
especially when most market participants are still using the 
traditional DM method.

This top-down method also produces a full set of risk measures. It 
is feasible to attribute the P\&L movement of a cash CLO trading book 
using the change of the average underlying loan prices and the 
index cash CLO tranche prices. Even though it is not feasible to hedge
CLO tranches by trading individual loans, it is certainly possible 
to macro hedges the risk of overall loan market movement based 
on the cash CLO deltas from the top-down model. If the market 
develops and certain ``index''
cash CLO deals becomes easier to short (for example, via TRS), 
then it is also possible to hedge the correlation risk of a
cash CLO book via the tranche01 risk from the model. Being able
to meaningfully define risk measures and devise their hedging 
strategies for a cash CLO book is certainly another big improvement 
over the DM based method.

This method is also computationally efficient since there is 
only a limited number of scenarios (Figure \ref{cadrs}) 
to run for each deal. The calibration, pricing and risk
measures of cash CLO tranches can be computed very efficiently 
using the the standard Intex tool, and this is no need to build
any custom cashflow waterfall engine. This top-down
method is very easy to implement and operate in practice as
most cash CLO market participants already use the Intex tool.
Using this top-down method, different market participants 
will reach the same CLO tranche prices if they can agree on 
a standard set of market scenarios like those listed 
in Figure \ref{cadrs}, and if they can establish a  
poll to determine the prices of a small set of representative
``index'' cash CLO tranches. Both of these two steps 
are well within reach therefore this method has the potential 
to bring much more pricing transparency to the cash CLO 
market.

\bibliographystyle{apsr}
\bibliography{creditref}

\end{document}